\documentclass[doublespacing]{elsart2}
\usepackage{bbm}
\usepackage{amsmath,amssymb,amsfonts,epsf}
\usepackage{graphicx}

\usepackage{lineno}

\usepackage{amssymb}

\begin{document}

\begin{frontmatter}



\title{BHP universality and gaussianity in sunspot numbers fluctuations}


\author[FEUP]{\corauthref{cor1}R. Gon\c calves}
\ead{rjasg@fe.up.pt}
\author[DMUM]{A. A. Pinto}

\corauth[cor1]{Universidade do Porto,
R. Dr. Roberto Frias,
4200-465 Porto Portugal,
tel/Fax: +351 225081707/1446}

\address[FEUP]{Faculdade de Engenharia\\
R. Dr. Roberto Frias, 4200 - 465, Porto, Portugal\\}

\address[DMUM]{Universidade do Minho\\
Campus de Gualtar, 4710 - 057 Braga, Portugal\\}

\begin{abstract}

We analyze the famous Wolf's sunspot numbers. We discovered that the distribution of the sunspot number fluctuations is
 a mixture of the BHP distribution with the Gaussian distribution.

\end{abstract}

\end{frontmatter}
\section{Introduction}
\label{sec:Int}

The sunspots are relatively dark areas on the surface of the sun caused by strong concentration of magnetic flux. 
Lu et al.  \cite{Luetal91} stated that the physical picture that arises, associated to sunspots, is that solar flares are avalanches of many reconnection events analogous to avalanches of sands in the models of Bak et al. \cite{BTW88}.  The relation between small-scale processes and the statistics of global flare properties, that follows from the self-organized magnetic field configuration, provides a way to learn about the physics of unobservable small-scale reconnection processes.
On the other hand, Bramwell et al. \cite{bramwell2000} showed a relation between self-organized criticality and the universal BHP distribution, named after the work of Bramwell, Holdsworth and Pinton \cite{BHP1998}.
We make a bridge between the two literatures, showing that the distribution of the sunspot numbers fluctuations is close to a mixture of the BHP with a Gaussian distribution.

\section{A BHP and a gaussian mixture fit of the sunspot numbers fluctuations}
\label{sec:BHPsunspots}

The average duration of the sunspot cycle is 133 months, but cycles as short as 9 years and as long as 14 years have been observed by Rabin et al. \cite{Rabinetal}.
Following Bramwell et al. \cite{bramwellfennelleuphys2002} and J\'anosi et al. \cite{ImreJason99}, we define the \emph{sunspot numbers mean period} $w_\mu(t)$ by
\begin{equation}
         w_\mu(t)=\frac{1}{T}\sum_{j=0}^{T-1} w(t+j*133) 
        \label{rjasgeq1}
\end{equation}
\noindent
and the \emph{sunspot numbers standard deviation period} $w_{\sigma}(t)$ by
\begin{equation}
         w_{\sigma}(t)=\sqrt{\frac{\sum_{j=0}^{T-1} w(t+j*133)^2}{T}-{w_\mu(t)}^2}\quad ,
        \label{rjasgeq2}
\end{equation}
\noindent
where $T=23$ is the number of observed cycles.
The \emph{sunspot numbers fluctuations} $w_f(t)$ is given by
\begin{equation}
          w_f(t)=\frac{w(t)-w_\mu(t)}{w_\sigma(t)}\quad.
        \label{rjasgeq3}
\end{equation}
\noindent

\begin{figure}[!htb]
\begin{center}
\includegraphics[width=10cm]{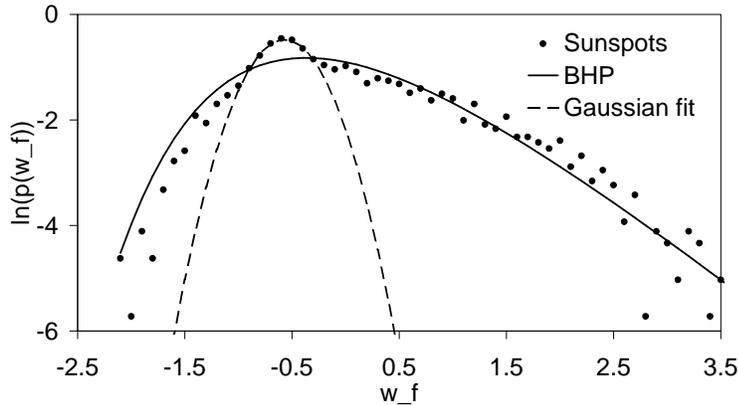}
\caption{\footnotesize{Histogram of the sunspot numbers fluctuations with a Gaussian and BHP fits, in a semi-log plot.}}
\label{fig:figure5}
\end{center}
\end{figure}

In Figure \ref{fig:figure5}, we show a fit of the BHP pdf to the histogram of the sunspot numbers fluctuations. 
For sunspot numbers fluctuations away of the mode value, the BHP pdf is close to the histogram.
In the range -0.9 to -0.3, close to the mode value of the histogram, we notice a higher concentration of the sunspot numbers fluctuations compared with the BHP pdf. Hence, we add a Gaussian fit to describe the concentration of the sunspot numbers fluctuations in this range.
We conclude that the histogram of the sunspot numbers fluctuations can be well approximated by a mixture of a Gaussian pdf  with the BHP pdf.

\section*{Acknowledgements}
We would like to thank Nico Stollenwerk for showing us the relevance of the BHP distribution.

\end{document}